
\font\mybb=msbm10 at 10pt

\def\bb#1{\hbox{\mybb#1}}

\def\bZ{\bb {Z}}

\def\bT{\bb {T}}
\def\bR{\bb {R}}
\def\bE{\bb {E}}
\def\bM{\bb {M}}

\def\bp{\rm p}
\def\bq{\rm q}


\tolerance=10000
\input phyzzx

 \def\unit{\hbox to 3.3pt{\hskip1.3pt \vrule height 7pt width .4pt \hskip.7pt
\vrule height 7.85pt width .4pt \kern-2.4pt
\hrulefill \kern-3pt
\raise 4pt\hbox{\char'40}}}

\REF\PKT{P.K. Townsend,  Phys. Lett. {\bf 350B}
(1995) 184; {\it p-Brane Democracy}, hep-th/9507048, to appear in
proceedings of the PASCOS/Hopkins workshop, March 1995.}
\REF\HT{C.M. Hull and P.K. Townsend, Nucl. Phys. {\bf B438} (1995) 109.}
\REF\sch{J.H. Schwarz, {\it String Theory Symmetries}, hep-th/9503127;
 {\it The
 power
of M-theory}, hep-th/9510086; {\it M-theory extensions of T-duality},
hep-th/9601077.}
\REF\sen {A.Sen Int. J. Mod. Phys. A {\bf 9} (1994) 3707.}
\REF\sensch{ J.H. Schwarz and A. Sen, Phys. Lett. {\bf 312B} (1993) 105.}
\REF\font{A. Font, L. Iba\~nez, D. L\"ust and F. Quevedo,
Phys. Lett. {249B}
 (1990)
35.}
\REF\witten{E. Witten, Nucl. Phys. {\bf B443} (1995) 85.}
\REF\duff{M.J. Duff, Nucl. Phys. {\bf B442} (1995) 47.}
\REF\cosmic{ B.R. Greene, A. Shapere, C. Vafa, and S.T. Yau,
Nucl. Phys.
 {\bf B337}
(1990) 1}
\REF\gg{G.W. Gibbons, M.B. Green and M.J. Perry, Phys. Lett. {\bf
370B}
 (1996) 37. }
\REF\plu{H. Lu, C.N. Pope, E. Sezgin and K.S. Stelle,  Phys. Lett. {\bf 371B}
(1996) 46.}
\REF\bgpt{ E. Bergshoeff, M.B. Green, G. Papadopoulos, M. de Roo and
 P.K. Townsend,
{\it Duality of type II 7-branes and 8-branes}, hep-th/9601150.}
\REF\pw{J. Polchinski and E. Witten, Nucl. Phys. {\bf B460} (1996) 525.}
\REF\wa{ E. Witten, Commun. Math. Phys. {\bf 80} (1981) 381.}
\REF\gh{G.W. Gibbons and C.M Hull,  Phys. Lett. {\bf 109B} (1982) 190.}
\REF\GHT{G.W. Gibbons, G.T. Horowitz and P.K. Townsend, Class. Quantum Grav.
{\bf 12} (1995) 297.}
\REF\lpt{J.M. Izquierdo, N.D. Lambert, G. Papadopoulos and P.K. Townsend, Nucl.
Phys.{\bf B460} (1996) 560.}
\REF\pht{J.M. Izquierdo, P.S. Howe, G. Papadopoulos and P.K. Townsend,
 Nucl. Phys.
{\bf B467} (1996) 183.}
\REF\Nep{R. Nepomechie, Phys. Rev. {\bf D31} (1984) 1921; C. Teitelboim, Phys.
Lett. {\bf B167} (1986) 69.}
\REF\garry {G.W. Gibbons, Nucl. Phys. {\bf B207} (1982) 337.\hfill\break
G.W. Gibbons and K. Maeda, Nucl. Phys. {\bf B298} (1988) 741.}
\REF\DL{M.J. Duff and J.X. Lu, Nucl. Phys. {\bf B416} (1994) 301.}
\REF\DFKR{M.J. Duff, S. Ferrara, R.R. Khuri and J. Rahmfeld,
Phys. Lett.
 {\bf 356B}
(1995) 479.}
\REF\BBO{E. Bergshoeff, H.J. Boonstra and T. Ort{\'{\i}}n, Phys. Rev. {\bf D53}
(1996) 7206.}
\REF\DLB{M.J. Duff and J.X. Lu, Phys. Lett. {\bf 273B} (1991) 409.}
\REF\doug{M. Douglas, {\it Branes within Branes}, hep-th/9512077.}
\REF\strom {A. Strominger, {\it Open p-branes}, hep-th/9512059.}
\REF\tsey {A.A. Tseytlin, {\it Selfduality of Born-Infeld action and Dirichlet
3-brane of IIB superstring theory}, hep-th/9602064.}
\REF\ggu{M.B. Green and M. Gutperle, {\it Comments on three-branes},
 hep-th/9602077.}
\REF\at{E.R.C. Abraham and P.K. Townsend, Nucl. Phys. {\bf B351} (1991) 313.}
\REF\ptc{G. Papadopoulos and P.K. Townsend, {\it Intersecting p-branes},
hep-th/9603087.}
\REF\SaSe{A. Salam and E. Sezgin, Nucl. Phys. {\bf B258} (1985) 284.}
\REF\Witb{E. Witten, Phys. Lett. {\bf 86B} (1979) 283.}
\REF\stelle{D.J. Duff and K.S. Stelle, Phys. Lett. {\bf 253B} (1991) 113.}
\REF\guven{R. G\"uven, Phys. Lett. {\bf 276B} (1992) 49.}
\REF\aat{A.A. Tseytlin, {\it Harmonic superpositions of M-branes},
 hep-th/9604035.}
\REF\hw {P. Horava and E. Witten, Nucl. Phys. {\bf B460} (1996) 506.}
\REF\lupo{ H. L\"u and C.N. Pope, Nucl. Phys. {\bf B465} (1996) 127.}
\REF\pol {J. Polchinski, S. Chaudhuri, C. V. Johnson, {\it Notes on D-branes},
hep-th/9602052.}


\Pubnum{ \vbox{ \hbox{R/96/25}  \hbox{}} }
\pubtype{}
\date{April, 1996}

\titlepage

\title { {\bf A brief guide to p-branes}\foot{Talk given at the workshop on
`STU-dualities and non-perturbative phenomena in superstring and supergravity'
 at CERN, 27 Nov-1 Dec 1995.}}

\author{G. Papadopoulos}
\address{DAMTP,
\break
University of Cambridge,
\break 
Silver Street, 
\break
Cambridge CB3 9EW,\break U.K.}

\abstract{We describe the qualitative properties of $p$-brane solutions of
supergravity theories and  present two examples of
$p$-brane solutions, first the dyonic membrane solutions of N=2 D=8
supergravity and second the intersecting M-brane solutions of D=11
 supergravity. }

\endpage


\chapter{Introduction}

The last two years have seen remarkable progress towards understanding the
non-perturbative properties of superstring theory.  This was achieved
by
 making a
number of conjectures about the non-perturbative symmetries, called duality
symmetries, of the superstring theories (see for example [\PKT-\duff]).  These
conjectures are a generalisation of the electromagnetic duality conjecture of
Montonen and Olive for Yang-Mills theories.  An essential role in understanding
the non-perturbative properties of superstring theories is played by
their 
associated
effective supergravity theories.  This is because first the duality 
symmetries of
superstring theory appear naturally at the level of supergravity
theories, 
in fact
they are discrete subgroups of the supergravity duality groups [\HT]. 
Second certain
classical solutions of supergravity theory, called `$p$-branes', 
are the analogue of
monopoles and dyons solutions of Yang-Mills theories. Therefore some 
$p$-brane
solutions of supergravity theories are associated with
non-perturbative 
states in
superstring theory.   We shall refer to $p$-branes with $-1\leq p\leq 2$ as
instantons ($p=-1$), particles ($p=0$), strings ($p=1$) and membranes 
($p=2$).  

In the first part of this paper, we will explain the qualitative properties of
$p$-brane solutions of supergravity theory.  These include the following: the
general ansatz that describes a $p$-brane solution of a D-dimensional 
supergravity
theory, a brief description of an energy bound in the context of 
$p$-branes, the
definition of the dual $\tilde p$-brane of a $p$-brane, the direct 
reduction and
wrapping of $p$-branes, and the intersection of $p$-branes.  In 
the second part of
the paper, we will give two examples of $p$-brane solutions, 
one will be the dyonic
membranes of D=8 N=2 supergravity theory and the other will 
be a class of solutions
of D=11 supergravity with the interpretation of intersecting 
membranes and 5-branes
(M-branes).


\chapter {Qualitative properties of $p$-branes}

\section{Supergravity}

The effective theory for the massless modes of a D-dimensional superstring
theory is a  D-dimensional supergravity theory.  The bosonic part of the
Lagrangian\foot{We have ignored potential $\alpha\prime$ corrections in the
supergravity Lagrangian and we have set $\alpha\prime=1 $\ .} of 
such supergravity
theory is
$$
{\cal L}={\cal L}(g,F,\phi)
\eqn\supone
$$
where $g$ is the spacetime metric, $F=\{F^\alpha; \alpha=1,\dots,m\}$ are
$(p_\alpha+2)$-form field strengths and $\phi$ are sigma model matter 
fields with
target space a $G/H$ coset.  The group $G$ acts on the scalars by
its standard left action on the coset $G/H$.  It also acts on the 
field strengths $F$
and their Poincar\'e duals with some representation in such way 
that the {\sl field
equations} of the supergravity theory are {\sl invariant}. 
(In the full theory the
fermions also transform under the action of the group $G$). 
In what follows, apart
from the bosonic sector of the the D-dimensional supergravity 
theory above, we shall
use the supersymmetry transformations of the gravitivi,
$\psi$, and the other spin $1/2$ fermions, $\lambda$, of the 
theory. Let $D$ be the
covariant derivative of the spin connection of the 
metric $g$. The supersymmetry transformations of $\psi$ and $\lambda$
evaluated at a background for which all fermions vanish takes the form
$$
\eqalign{
\delta\psi&\equiv \hat D\eta =D(g,\omega)\eta+T(g, F,\phi)\eta
\cr
\delta\lambda&=L(g,\phi,F)\eta\ ,}
\eqn\suponea
$$
where $\eta$ are the supersymmetry parameters, and  $T$, $L$ are  
matrices that
depend upon the  fields of the theory
\foot{We have suppressed space-time indices, spinor indices 
and indices that count
the number of supersymmetries.}. The action of the group $G$ 
on the various fields of
the supergravity theory induces an action on the supersymmetry
transformations \suponea. Under this action, the equations \suponea\
transform in some representation of $G$ provided that the 
supersymmetry parameters
$\eta$ are transformed in a suitable way. This fact will be used in section 2.3
to argue for the invariance of an energy bound under the action 
of the group $G$.  

The vacua of the above supergravity theory are parameterised 
by the $G/H$ coset. 
Using the Dirac quantisation for the charges of the $p$-brane
solutions of \supone, one can argue that the symmetry $G$ of the field
equations of the theory is quantum mechanically broken to the discrete subgroup
$G(\bZ)$.  The U-duality conjecture then involves the assertion 
that $G(\bZ)$ is a
symmetry of the full associated superstring theory [\HT]. The 
vacua  of the theory
that lie in the orbits of $G(\bZ)$ acting on $G/H$ are identified, i.e. quantum
mechanically the space of vacua is $G(\bZ)\\G/H$.

\section {$p$-branes} 

We are seeking solutions of supergravity theories that have the interpretation
of parallel infinite planar p-dimensional extended objects located in a
D-dimensional spacetime, i.e. the spacetime is foliated with parallel 
leaves that
are isometric to (p+1)-dimensional Minkowski spacetime.  We shall refer to such
solutions as `$p$-branes'. We shall also require that such solutions  
should have a
well defined notion of mass $M$ and charge $\bp$ per unit volume, and that they
should be extreme, i.e.
$M=|\bp|$. Since the $p$-brane carries
charge $\bp$, it couples  naturally to a $(p+1)$-form gauge potential $A$. 
Note that apart from the non-trivial spacetime metric $g$ and 
$(p+2)$-form field
strength
$F=dA$ which are necessary to describe a $p$-brane solution, the
$p$-brane solution may include some non-constant scalar fields $\phi$.

The above description of the $p$-brane solutions of a supergravity 
theory requires
that they must have a  $(p+1)$-dimensional Poincar\'e invariance;  
the Poincar\'e
group acts  on the co-ordinates $\{x^\mu; \mu=0,\dots, p\}$ which 
are identified
with the worldvolume co-ordinates of the 
$p$-brane.  Let $\{y^i;i=1, \dots, D_\perp\equiv D-p-1\}$ be the transverse
co-ordinates to the $p$-brane in the D-dimensional spacetime.  Then we have 
$$
D=D_\perp +p+1\ .
\eqn\form
$$
An ansatz for a $p$-brane solution of a supergravity theory is the following:
$$\eqalign{
ds^2&=A^2(y) ds^2(\bM^{p+1}) +B^2(y) ds^2(\bE^{D_\perp}) ,
\cr
F&=\epsilon_{(p+1)}\wedge dC(y)+\xi(y) \ ,
\cr
\phi&=\varphi(y)\ , }
\eqn\suptwo
$$
where  $ds^2(\bM^{p+1})$ is the  Minkowski metric, $ds^2(\bE^{D_\perp})$ is the
Euclidean metric, $\epsilon_{(p+1)}$ is the volume form on $\bM^{p+1}$, $A$, $B$ and
$C$ are functions of the transverse co-ordinates $y$ and $\xi$ is a closed
$(p+2)$-form on the transverse space.\foot{The part of the metric that
involves the
function $B$, as stated, does not follow from the requirements that we
have imposed on
the $p$-brane solutions but it turns out that the known $p$-brane solutions are
always of this form.} 

The ansatz \suptwo\ involves several unspecified functions that are
determined by solving the Killing spinors equations, 
$$
\eqalign{
 \hat D\eta&=0
\cr
L(e,\phi,F)\eta&=0\ ,}
\eqn\pthree
$$
and some of the field equations.
The Killing spinor equations  are the vanishing conditions of the
supersymmetry transformations \suponea\ and as we shall explain in 
the next section
are necessary conditions for the $p$-brane solutions to be extreme. The
$p$-brane solutions are then determined in terms of harmonic functions on
$\bE^{D_\perp}$; the unknown functions of the metric $ds^2$ and scalar
fields $\phi$
in \suptwo\ are usually expressed in terms of powers of harmonic 
functions, and the
field strength
$F$ is expressed in terms of the same harmonic functions and their
derivatives.  The harmonic functions on Euclidean spaces have point 
singularities
that are interpreted as the positions of the $p$-branes.

The asymptotic behaviour of the metric of a $p$-brane solution as
$|y|\rightarrow
\infty$ depends upon the dimension $D_\perp$ of the transverse space to the
$p$-brane.   If $D_\perp>2$, it can be arranged that the spacetime is 
asymptotically
Minkowski. This is due to the fact that  the harmonic functions on
$\bE^{D_\perp}$ for $D_\perp>2$  can be chosen  such that they approach to a
constant as $|y|\rightarrow \infty$.  For $D_\perp\leq 2$ the
spacetime 
exhibits 
different asymptotic behaviour. In the $D_\perp=2$ case, either the 
spacetime is
asymptotically
$\bM^{p+1}\times Y$, where $Y$ is a two-dimensional conical space, 
or the metric
exhibits logarithmic behaviour as $|y|\rightarrow \infty$ 
[\cosmic,\gg, \plu]. For
$D_\perp=1$, the $(D-2)$-brane separates the D-dimensional spacetime into two
asymptotic regions like a domain wall [\bgpt,\pw,\plu].  Finally $(D-1)$-branes
solutions are identified with the Minkowski vacuum of the 
associated supergravity
theory.  The behaviour of the metric at the positions of the 
$p$-branes is also of
interest. The metric is singular at those positions  but in
some cases these singularities are merely co-ordinate singularities.  
In addition,
if the relevant solution has non-constant scalar fields, then the nature of the
singularity at the positions of the $p$-branes depends upon the choice
of the frame
for the metric.   Nevertheless one can deduce important 
information by studying the
singularity structure of the $p$-branes at their positions. For example
$p$-brane solutions with time-like naked singularities 
(in their natural frame) are
usually interpreted as fundamental objects while non-singular solutions are
interpreted as $p$-brane monopoles and dyons.

\section{Energy per unit volume bound}

As we have already mentioned in the previous section, the $p$-brane 
solutions of
supergravity theories saturate an energy per unit volume bound.  Here we shall 
describe how such a bound can be derived (see refs [\wa-\lpt]). 
For this we define
the modified Nester tensor
$$
\hat E^{MN}={1\over 2}\bar\kappa \Gamma^{MNR}\hat D_R\kappa+ c.c
\eqn\gponea
$$
where $\kappa$ is a commuting spinor and $\hat D$ is the operator 
that appears in
the gravitini supersymmetry transformation law \suponea. Then one can show that
$$
D_N \hat E^{MN}= {\bar{\hat D}}_N \kappa \Gamma^{MNR}{\hat D}_R 
\kappa-{1\over2}
\bar\chi\Gamma^M\chi
\eqn\gponeb
$$
where $\chi$ are proportional to the expressions that appear in the 
supersymmetry
transformations \suponea\ of the fermions, $\lambda$ (with $\eta$, 
the supersymmetry
parameter, replaced by $\kappa$).

Next we shall consider a spacetime that asymptotically describes an 
extended object
as we have discussed in the previous section, i.e. one can distinguish the
directions along an extended object and the directions transverse to
it.  
Therefore
the spacetime is asymptotically
$\bM^{p+1}\times \bE^{D_\perp}$ along directions transverse to the
extended object. (We take here $D_\perp\geq 3$, the case for which 
$D_\perp\leq 2$
will be discussed at the end of the section).  Next we assume that 
the $p$-brane is
wrapped around a large torus,
$\bT^p$, so the topology of the spacetime at spatial infinity along 
the transverse
directions (`transverse spatial infinity') is  $\bT^p\times S^{D_\perp-1}$.
Using this and appropriate decay condition for the other fields, 
we can define the
transverse $D_\perp$ momentum $P$ and charge $\bq$ per unit volume of such
spacetime as follows:
$$
\bar\kappa_\infty \Gamma\cdot P \kappa_\infty \equiv {1\over 2 V_p
\Omega_{D_\perp-1}}\int_\infty dS^{MN} E_{MN}={1\over 2
\Omega_{D_\perp-1}}\int_{S^{D_\perp-1}} dS^{ij} E_{ij}
\eqn\gpone
$$
and 
$$
\bq\equiv {1\over V_p \Omega_{D_\perp-1}}\int_\infty \tilde F={1\over
\Omega_{D_\perp-1}}\int_{S^{D_\perp-1}} \tilde F\ ,
\eqn\gptwo
$$
where $E$ is the Nester tensor associated with the spin connection 
of the metric,
$\kappa_\infty$ is the asymptotic value of the spinor
$\kappa$, 
$V_p$ is the volume of the unit p-torus,
$\Omega_{D_\perp-1}$ is the volume of $(D_\perp-1)$-sphere with unit radius and
$\tilde F$ is the Hodge dual\foot{The definition of the `electric'
charge involves the electro-magnetic dual field strength $G$ of $F$ 
rather than its
Hodge dual $\tilde F$ that we have used here for simplicity; see for example
section 3.} of $F$.   Then, one can show that
$$
\bar\kappa_\infty K\kappa_\infty={1\over V_p \Omega_{D_\perp-1}}
\int dS^{MN}\hat
E_{MN}\geq 0
\eqn\gpthree
$$
where $K$ is
$$
K=\Gamma\cdot P+ Z(  \langle\phi\rangle, \bq)\ ,
\eqn\gpfour
$$
and  $\langle\phi\rangle$ are the asymptotic values of the scalars $\phi$.  The
equality in \gpthree\ can be established by following the definition 
of the various
quantities  and $Z$ can be easily determined from the modified 
Nester tensor.  To
show the inequality in \gpthree\ is more involved; for this one 
makes use of the
\gponeb\ and the `Witten condition' [\wa]. Finally from \gpfour\ 
and \gpthree\ one
can establish a bound for the mass per unit volume of the 
spacetime in terms of its
charges per unit volume.  The invariance of the bound under 
the $G$ group action is
an immediate consequence of the invariance of the Nester tensor
$\hat E$.  The latter follows from the transformation properties of the 
supersymmetry transformations \suponea\ under the action of the group $G$.  
It is clear from \gponeb\ that a configuration saturates the 
bound  provided that it
satisfies the Killing spinor equations \pthree\ (with $\eta$, the supersymmetry
parameter, replaced by $\kappa$).

Now we turn to examine the case where $D_\perp\leq 2$. 
For $D_\perp=2$, many steps
in the proof of the bound remain the same.  However if one assumes that
the asymptotic behaviour of the  spacetime along the transverse directions  is 
$\bR\times
\bT^{D-3}\times Y$ where
$Y$ is a two-dimensional conical space, then for all 
configurations that satisfy the
Witten condition the bound is always saturated.  This follows from a slight
modification of the proof of an energy theorem in 
$2+1$ dimensions given in [\pht]. 
The $D_\perp=1$ case involves a spacetime which is topologically
$\bR\times \bT^{D-2}\times \bR$ as one approaches the transverse spatial
infinity.  The transverse spatial infinity in this case is just the
disjoint union of two
$\bT^{D-2}$ and the calculation of the mass and charge 
per volume simply involves the
evaluation of the corresponding expressions at two points at infinity.

\section{Dual and dyonic $p$-branes}

The dual of a $p$-brane coupled to a (p+1)-form
potential, $A$, in a D-dimensional spacetime, is a
$\tilde p$-brane coupled to a $(\tilde p+1)$-form potential, $\tilde A$,
such that the form field strength $G=d\tilde A$ has rank $D-p-2$  and it
is related to the Poincar\'e dual of the $(p+2)$-form field 
strength $F=dA$\foot{The 
precise definition of $G$ can be somewhat complicated and it
depends on the way that Bianchi identities and field equations are
interchanged under electric-magnetic duality.}.  It follows 
from this definition that
[\Nep]
$$
\tilde p=D-p-4\ .
\eqn\gpfour
$$
It is clear that if the $p$-brane carries `electric' charge 
$\bq$, its dual $\tilde
p$-brane carries `magnetic' charge $\bp$, where
$$
\bp={1\over \Omega_{\tilde D_\perp-1}}\int_{S^{\tilde D_\perp-1}} F\ ;
\eqn\gpfive
$$
$\tilde D_\perp$ is the number of transverse directions to the $\tilde
p$-brane. 

The possibility arises for dyonic $p$-branes, i.e. those that carry
both electric and magnetic charges.  For such $p$-branes, 
$p=\tilde p$, and from
\gpfour,  we get
$$
p={D\over 2}-2\ .
\eqn\gpsix
$$
So one gets dyons in D=4 [\garry], dyonic strings in D=6 [\DL,\DFKR], dyonic
membranes in D=8 [\lpt, \BBO], and finally dyonic 3-branes 
in ten dimensions [\DLB].

\section{Direct reduction and wrapping of $p$-branes}

Given $p$-brane solutions of a supergravity theory in $D$ dimensions,  one can
construct new $q$-brane solutions in $D-k$ dimensions using `direct reduction' and
`wrapping' of $p$-branes.  The first procedure involves 
Kaluza-Klein (KK) reduction
in $k$ directions on the co-ordinates of the transverse space. 
Consistency with KK
ansatz requires that the harmonic functions should be independent from these
directions. With this procedure one gets a $p$-brane in $D-k$
dimensions. 
On the
other hand wrapping involves KK reduction on $k$ space-like directions of the
worldvolume of the $p$-brane.  The resulting solution is a
$(p-k)$-brane in $D-k$ dimensions.  It is also possible to employ simultaneous
direct reduction and wrapping.  In which case if one wraps a $p$-brane in $k_1$
directions and directly reduces it in $k_2$ directions, the result is a
$(p-k_1)$-brane in $d-k$ dimensions, $k=k_1+k_2$.  The above reduction methods
for $p$-branes are particularly simple to perform in the context torus
compactifications of supergravity theories but they can also be 
extended to other
compactifications of supergravity theories, like $K_3$ and Calabi-Yau ones. The
wrapping of $p$-branes, in the latter case,  is around the various
co-homology co-cycles of the compactifying space.  The direction of 
direct reduction
and wrapping of
$p$-branes can be reversed.  Consequently one can lift $p$-brane solutions of
supergravity theories from lower dimensions to higher ones.

\section{Intersecting $p$-branes}

So far we have studied solutions of supergravity theories that have the
interpretation as parallel $p$-branes in a D-dimensional spacetime.  There are,
however, solutions of supergravity theories with somewhat different 
interpretation as
intersections of two or more orthogonal $p$-branes. The novelty is that such
configurations may be extreme and preserve some of the 
supersymmetry\foot{ They may
though exhibit different asymptotic behaviour from that of 
parallel $p$-branes.}. 
There are many ways to approach this topic [\doug-\at]; here we will follow 
[\ptc].  Consider the orthogonal intersection of
$p_\alpha$-branes,
$\alpha=1,\dots,L$, on a $r$-brane in D dimensions.  The
Poincar\'e invariance of the worldvolume of all $L$ orthogonal 
$p_\alpha$-branes
will be broken to that of the $r$-brane that lies in the intersection.
Therefore
the corresponding solutions  will have
$(r+1)$-dimensional Poincar\'e invariance.  The normal bundle of
the $r$-brane imbedded in $(D-1)$-dimensional space can be decomposed into 
$\ell$ directions along the tangent bundles of the $p_\alpha$-branes, 
and into the
remaining $D_\perp$ directions\foot{We have used the same symbol, $D_\perp$, to
denote both the number of transverse directions of a $p$-brane 
and the number of
overall transverse directions of $L$ intersecting $p_\alpha$-branes. 
The distinction
between the two uses of it will be clear from the context.}.  
We shall refer to the
former as the `relative' transverse directions and to the 
latter as the `overall'
transverse directions.  It is clear that
$$
D=r+\ell+D_\perp+1\ .
\eqn\intone
$$
Let $\{x^\mu; \mu=0,\dots, r\}$ be the worldvolume coordinates 
of the $r$-brane,
$\{u^a; a=1,\dots,
\ell\}$ be the co-ordinates along the relative transverse 
directions and $\{y^i,
i=1, \dots, D_\perp\}$ be the co-ordinates along the overall transverse directions.
Then the ansatz of a spacetime metric that describes $L$
intersecting $p_\alpha$-branes, $\alpha=1,\dots,L$, is
as follows:
$$
ds^2=A^2(u,y)ds^2(\bM^{r+1}) +B_{ab}(u,y) du^a du^b+ C_{ij}(u,y)  dy^i dy^j\ .
\eqn\suptwo
$$
The functions  $A,B,C$  are
determined by solving killing spinor and field equations. It is expected\foot{
In the limiting case of the intersection of a membrane and a $5$-brane 
with one to include the other (see section 4), the proportion of 
the supersymmetry
preserved is
$1/2$.}  that if such solution preserves some supersymmetry the proportion of
supersymmetry preserved is $1/2^L$.

To find the (magnetic) dual configuration in D dimensions of  $L$
orthogonal $p_\alpha$-branes intersecting on a $r$-brane with 
number of relative
transverse directions
$\ell$, we wrap the
configuration  to $d\equiv (D-\ell)$-dimensions along the relative transverse
directions. In
$d$ dimensions the solution  becomes an $r$-brane solution 
and from \gpfour\  its
magnetic dual is a $\tilde r$-brane with
$$
{\tilde r}=d-r-4\ .
\eqn\inttwo
$$
Now the magnetic dual configuration of the $L$ orthogonal
$p_\lambda$-branes intersecting on a $r$-brane (in D dimensions) 
should reduce to
the ${\tilde r}$-brane after wrapping along $\ell$ relative 
transverse directions to
$d$ dimensions.  This implies that
$$
D={\tilde r}+\ell+{\tilde D}_\perp\ +1 ,
\eqn\intthree
$$
where ${\tilde D}_\perp$ is the number of overall transverse directions of the
magnetic dual configuration.  The equations \inttwo\ and 
\intthree\ can be solved
for ${\tilde r}$ and ${\tilde D}_\perp$ as follows:
$$
\eqalign{
{\tilde r}&=D-\ell-r-4
\cr
{\tilde D}_\perp&=r+3 \ .}
\eqn\intfour
$$
Note that the equations which can be obtained from \intfour\ by exchanging 
$(r, D_\perp)$ with $(\tilde r, \tilde D_\perp)$ are also valid.


\chapter{Dyonic membranes}

As we have already mentioned  membranes in eight dimensions 
($D=8$) can carry both
electric and magnetic charges. The possibility then arises for the existence of
dyonic membrane solutions in a D=8 supergravity.  Such solutions 
have already been
found in [\lpt] in the context 
of N=2 D=8 supergravity [\SaSe]. The relevant Lagrangian is a
consistent truncation of the D=8 supergravity  Lagrangian with 
fields the metric
$g$, two scalars $\sigma$ and
$\rho$ and a 4-form field strength $F$.  The truncated Lagrangian 
is as follows:
$$
\eqalign{
{\cal L} = N\Bigg\{\sqrt{-g}\big[ &R - 2\partial_\mu \sigma\partial^\mu\sigma -
2e^{4\sigma}\partial_\mu \rho\partial^\mu\rho  -{1\over
12}e^{-2\sigma}F_{\alpha\beta\gamma\delta}F^{\alpha\beta\gamma\delta}\big] \cr
&-{1\over 144}\varepsilon^{\mu\nu\rho\sigma\alpha\beta\gamma\delta}\rho 
F_{\mu\nu\rho\sigma}F_{\alpha\beta\gamma\delta} \Bigg\}\ ,}
\eqn\onec
$$
for some normalisation factor $N$.
The scalar fields $\sigma$ and $\rho$ take values in the coset 
$SL(2,\bR)/U(1)$ and
it is convenient to define the fields
$$\eqalign{
\lambda&=2\rho+i e^{-2\sigma}
\cr
G&=e^{-\sigma} \tilde F-2\rho F
\ , }
\eqn\oned
$$
where $\tilde F$ is the Poincar\'e dual of $F$.  The new field 
$\lambda$ transforms
with fractional linear transformations under $SL(2,\bR)$ while the pair $(F,G)$
transforms as a doublet under the same group.  The field 
equations of this theory
are invariant under this action of $SL(2,\bR)$.  

We shall be interested in membrane solutions of the equations of
motion of \onec\ that are asymptotically flat as one approaches 
spatial infinity 
in transverse directions. The transverse spatial infinity in this case is
topologically $S^4\times \bR^2$. The dyonic membrane solutions can be
characterised by their magnetic, $\bp$, and electric, $\bq$, charges.  These
charges, after an appropriate choice for the normalisation constant $N$ of the
Lagrangian and for the unit of electric charge, are
$$
\bp= {1\over \Omega_4}\int_{S^4}\! F\ , \qquad \bq 
= {1\over \Omega_4}\int_{S^4}\!
G\ ,
\eqn\chargetwo
$$
where the integral is over a 4-sphere cross-section of transverse spatial
infinity and $\Omega_4$ is the volume of the 4-sphere with unit radius.  The
dyonic membrane solutions of the field equations of
\onec\ are [\lpt]
$$
\eqalign{
ds^2 &= H^{-{1\over2}}ds^2(\bM^3) + H^{1\over2} ds^2(\bE^5) \cr
F &= {1\over2}e^{\langle\sigma\rangle}\Big( \cos\psi \star dH + \sin\psi\;
dH^{-1} \wedge \epsilon_3\Big)
\cr
\lambda &= 2\langle \rho\rangle + e^{-2\langle\sigma\rangle} \cdot {(1-H)\sin
2\psi + 2i H^{1\over2}\over 2(H\cos^2\psi + \sin^2\psi)}
\ ,}
\eqn\dyonsthree
$$
where star is the
Hodge star in $\bE^5$, 
$H$ is a harmonic function on $\bE^5$, $\psi$, $\langle\rho\rangle$ 
and $\langle
\sigma\rangle$ are  parameters of the solution, and $\epsilon_3$ is the volume
form in $\bM^3$.  As
$|y|\rightarrow \infty$, $y\in\bE^5$, the metric approaches that of Minkowski
spacetime and the field $\lambda$ approaches the vacuum  
$$
\langle \lambda\rangle=2 \langle \rho\rangle +i e^{-2 \langle
\sigma\rangle}\ .
\eqn\dyonfour
$$
The solutions \dyonsthree\ preserve half the supersymmetry. 
The purely `electric' solution in the vacuum $\langle \lambda\rangle=i$ can be
identified as that for which $\cos\psi=0$ in which case the 
purely `magnetic' one is
that with $\sin\psi=0$.

Following the steps described in the previous section, we can derive the
bound
$$
M^2\geq {1\over4} \bigg[ e^{2\langle \sigma\rangle}(\bq+2\langle \rho\rangle
\bp)^2+e^{-2\langle \sigma\rangle}\bp^2\bigg]\ ,
\eqn\dyonseven
$$
on the mass per unit volume, $M$;  for this we have used
$$
\hat D\kappa=D\kappa-{1\over2}\Gamma_9\kappa e^{2\sigma} d\rho+{1\over 96}
\Gamma^{\alpha\beta\gamma\delta}\Gamma\kappa e^{-\sigma}
F_{\alpha\beta\gamma\delta}\ .
\eqn\dyonfive
$$
This bound is $SL(2,\bR)$ invariant and the dyonic membrane 
solutions \dyonsthree\
saturate this bound as it is easy to show by direct computation.

Dyonic membranes have similar quantum properties to those of dyons.  Indeed
combining the Nepomechie-Teitelboim (N-T) quantization condition for extended
objects with  Schwinger-Zwanziger quantization condition for dyons, one can
find an analogue of the latter that applies in the 
context of dyonic $p$-branes.
This generalized N-T quantization condition for two dyonic 
membranes with charges
$(\bp,\bq)$ and $(\bp',\bq')$ takes the simple (manifestly
$Sl(2;\bR)$ invariant) form
$$
\bq\bp'-\bq'\bp \ \in \ \bZ\ .
\eqn\dirac
$$
As for dyons in D=4 [\Witb], this formula allows fractional $\bq$ for dyonic
membranes.  

The duality group of the D=8 N=2 supergravity is 
$G=SL(3;\bR)\times SL(2;\bR)$.  The
4-form field strength $F$, and its dual, transform under the 
$({\bf 1}, {\bf 2})$
representation of the supergravity duality group. The 
U-duality group in this case is
$G(\bZ)=SL(3;\bZ)\times SL(2;\bZ)$ and includes the 
T-duality group $SO(2,2; \bZ)$ of
D=8 type II superstring as a proper subgroup.  Despite the fact that
T-duality is a perturbative symmetry in the context of superstring
theory, it acts
on $F$ via a generalised electromagnetic duality transformation.  
For a more detail
discussion of the applications of dyonic membranes in superstring theory 
see [\lpt].

\chapter{Intersecting M-branes}

As we have mentioned, certain solutions of supergravity theories have
the interpretation of intersecting $p$-branes.  Examples of such solutions can 
be found in  D=11 supergravity.  The D=11 supergravity has 
bosonic field content a
metric $g_{(11)}$ and a 4-form field strength $F_{(11)}$.  
There are two  well known
$p$-brane solutions of D=11 supergravity the membrane [\stelle] and 5-brane
[\guven].   We will refer to them collectively as M-branes.  
However  apart from
these two solutions, D=11 supergravity has additional solutions  that solve the
Killing spinor equations.  A class of such solutions given in [\guven] is the
following:
$$ 
\eqalign{
ds^2_{(11)}&=- H^{-{2n\over 3}} dt^2+H^{n-3\over3} ds^2(\bE^{2n})+H^{n\over3}
ds^2(\bE^{10-2n})
\cr
F_{(11)}&=-3 dt\wedge dH^{-1}\wedge J\ ,}
\eqn\elone
$$ 
where $J$ is a `constant' complex structure in $\bE^{2n}$ and $n=1,2,3$. For
$n=1$, we get the membrane solution of D=11 supergravity [\stelle].  
The remaining
two solutions ($n=2,3$) can be thought as the geometric intersections 
of two and
three membranes at a $0$-brane correspondingly [\ptc]; the relative 
transverse space
is
$\bE^{2n}$ and the overall transverse space is $\bE^{10-2n}$.  
The proportion of
the supersymmetry preserved by the solutions \elone\ is $1/2^n$.

Using this interpretation for the solutions \elone\ and the general 
arguments about
intersecting $p$-branes in section 2.6,  one can immediately deduce that their
magnetic duals are again intersecting $p$-branes with
$$
\eqalign{
\tilde r&=7-2n
\cr
\tilde D_\perp&=3\ .}
\eqn\eltwo
$$
So for $n=1$ one gets precisely the 5-brane which is known to 
be the magnetic dual of
membrane.  The remaining cases, $n=2,3$, can be thought of as two 5-branes
intersecting at a 3-brane and three 5-branes intersecting at a string,
respectively.  Indeed the D=11 supergravity solutions that are 
the magnetic duals
of \elone\ are the following [\ptc]:
$$
\eqalign{
ds^2_{(11)}&=- H^{-{n\over 3}} ds^2(\bM^{8-2n})+H^{-{n-3\over3}}
ds^2(\bE^{2n})+H^{2n\over3} ds^2(\bE^3)
\cr
F_{(11)}&=\pm 3 \star dH\wedge J\ ,}
\eqn\elthree
$$
where $H$ is a harmonic function of $\bE^3$, star is the Hodge 
star in $\bE^3$ and
$n=1,2,3$. For $n=1$, one gets a special case of the 5-brane solution of D=11
supergravity.  (In the 5-brane solution of D=11 supergravity the 
harmonic function
$H$ is a function of $\bE^5$.) The remaining two solutions are 
the magnetic duals
of the intersecting membrane solutions \elone\ as they have 
been described above.
The proportion of the supersymmetry preserved by the solutions \elthree\ is
$1/2^n$.  Another solution of D=11 supergravity can be found
by lifting the dyonic membrane solutions of the previous section to D=11.
The D=11 interpretation of the lifted solution is of a membrane lying within a
5-brane [\lpt] and preserve $1/2$ of the supersymmetry.  Finally, 
there are three
more known $p$-brane-like solutions of D=11 supergravity the 
following [\aat]: (i) a
membrane and 5-brane intersecting at a string (preserving
$1/4$ of the supersymmetry), (ii) two membranes and a 5-brane intersecting at
particle  (preserving $1/8$ of the supersymmetry), and (iii)
a membrane and two 5-branes intersecting at a particle (preserving $1/8$ of the
supersymmetry). 

To study the solutions that one gets in D=10  from reducing the $p$-brane-like
solutions of D=11 supergravity, we denote with
$(r|p_1,\dots,p_L)$ the solutions of a supergravity theory that have the
interpretation of $L$ intersecting $p_\alpha$-branes, $\alpha=1,
\dots, L$, with common intersection a $r$-brane.  For example 
$(2|0)$ will denote a
membrane solution while $(0|2,2)$ will denote two intersecting membranes at a
$0$-brane.  There are three different ways to reduce a intersecting $p$-brane
solution the following: (i) wrapping the solution along the common intersecting
$r$-brane, (ii) wrapping the solution along one of the relative transverse
directions and (iii) directly reducing the solution along one of the overall
transverse directions.  Using the above notation, we denote the  $p$-brane-like
solutions of D=11 supergravity as follows: $(2|0)$, $(5|0)$, 
$(2|2,5)$, $(0|2,2)$,
$(3|5,5)$, $(1|2,5)$, $(0|2,2,2)$, $(1|5,5,5)$, $(0|2,2,5)$ and 
$(0|2,5,5)$ . Direct
reduction on $S^1$ along the overall transverse directions will 
produce solutions in
D=10 with exactly the same interpretation as the $p$-brane-like 
solutions in D=11. On
the other hand the wrapping of solutions along one relative 
transverse direction does
not apply to the membrane and 5-brane solutions but the remaining 
solutions reduce
to D=10 as follows: 
$(2|2,4)$, $(0|1,2)$, $(3|4,5)$, $(1|1,5)$, $(1|2,4)$ $(0|1,2,2)$, ,
$(1|4,5,5)$, $(0|1,2,5)$, $(0|2,2,4)$,  $(0|1,5,5)$ and $(0|2,4,5)$.  
Finally the
wrapping of solutions
 along the $r$-brane lying in the
common intersection does not apply to the D=11 M-brane solutions 
intersecting at a
particle but the remaining solutions reduce to D=10 as follows: $(1|0)$,
$(4|0)$, 
$(1|1,4)$, $(2|4,4)$, $(0|1,4)$ and
$(0|4,4,4)$.  It is worth mentioning that the
 $p$-brane-like solutions of D=11 supergravity reduce to $a=\sqrt 3, 1,
1/\sqrt 3$ (electric and magnetic) black hole solutions of D=4 N=8 supergravity
[\ptc]. The $a=0$ D=4 black-hole also has a D=11 interpretation 
but the relevant D=11
solution involves KK vectors [\ptc, \aat].


\chapter{Concluding Remarks}

We have presented the qualitative properties of solutions of 
supergravity theories
that  saturate a certain  bound and consequently satisfy a 
Killing spinor equation. 
Such solutions have a $p$-brane interpretation and are the 
analogues of monopoles and
dyons of Yang-Mills theories in the context of supergravity theories. We have
also briefly explained the use of the  $p$-brane solutions of supergravity
theories in the study of non-perturbative  superstring theory.  The
qualitative properties of
$p$-brane solutions that  have been described are the following: 
the general ansatz
for constructing such solutions, an energy bound, the `magnetic' dual $\tilde
p$-brane  of a $p$-brane, the direct reduction and wrapping of $p$-branes and
finally the intersection of $p$-branes.

Some $p$-brane solutions of $D<11$ supergravity theories can be obtained
from the $p$-brane-like solutions of D=11
supergravity using direct reduction and wrapping (there are
no supergravity theories in  $D>11$).   It is therefore of interest to
understand the $p$-brane-like solutions of the D=11 supergravity theory.  In
addition,  D=11 supergravity is the effective theory of a conjectured M-theory;
M-theory  upon  $S^1$ reduction to D=10 gives the type IIA superstring
[\PKT,\witten], and upon $S^1/\bZ_2$ reduction to D=10 gives the 
$E_8\times E_8$
heterotic string [\hw]. The solutions of D=11 supergravity that 
have been found, so
far, with a $p$-brane-like interpretation are the membrane and 
5-brane solutions
[\stelle, \guven] and their intersections [\guven,\ptc, \aat].  
The latter include
the intersection of two and three membranes at a particle,  
the intersection of a
membrane and a 5-brane at a string, the intersection of 
two membranes and a 5-brane
at a particle, the intersection of a membrane and two 
5-branes at particle, and the
intersection of two and three 5-branes at a 3-brane and 
a string, respectively. 
There is also another solution that has the interpretation of a membrane lying
within a 5-brane [\lpt]. It may be that this is not the whole story.  
For example,
there has been a classification of all
$p$-brane solutions in maximal supergravities [\lupo].  These solutions  can be
lifted and may produce new
$p$-brane-like  solutions in D=11.  Although the emphasis in this paper  was on
$p$-brane solutions of supergravity theories, a certain class of 
$p$-branes those
that couple to Ramond-Ramond fields may have an explanation 
within the context of
superstring theory as D-branes.  For this, one should allow 
strings with Dirichlet
boundary conditions in $D-p-1$ space-like directions in spacetime; the $p+1$
directions span the worldvolume of a D-brane (for a review see [\pol]). It is
likely that intersecting D-brane configurations are 
related to some $p$-brane-like
configurations of D=11 supergravity.

  
\vskip 1cm
\noindent{\bf Acknowledgements:} I would like to thank E. Bergshoeff,
G.W. Gibbons,
M.B Green, P.K. Townsend and A. Tseytlin for many helpful discussions.  I was
supported by a Royal Society University  Research Fellowship. 

\refout

\bye